\documentclass[twoside,12pt]{article}
\usepackage{epsfig}

\def\Journal#1#2#3#4{{#1} {#2} (#4) #3 }

\newcommand{\be}{\begin{equation}}
\newcommand{\ee}{\end{equation}}
\newcommand{\bea}{\begin{eqnarray}}
\newcommand{\eea}{\end{eqnarray}}

\begin{document}

\title{ \vspace{1cm} In-medium Properties of $\eta^\prime$ Meson}
\author{M.\ Nanova,$^{1}$ \\
on behalf of the  CBELSA/TAPS Collaboration\\
$^1$ II. Physikalisches Institut\\
Justus Liebig Universit\"at Giessen, Germany}

\maketitle
\begin{abstract}
Using the  Crystal Barrel(CB)/TAPS detector system at the ELSA accelerator facility in Bonn the $\eta^\prime$ photoproduction off nuclei (C, Ca, Nb and Pb) was studied via the hadronic decay channel $\eta^{'}\rightarrow \pi^{0}\pi^{0}\eta$ . Recent results on the in-medium properties of the $\eta^\prime$-meson, derived from the transparency ratio measurements, are presented. The absorption of the $\eta^\prime$-meson in nuclear matter is compared to the properties of other mesons ($\eta$ and $\omega$).

\end{abstract}
\section{Introduction}
The properties of $\eta^\prime $ are largely governed by the dynamics of the QCD $U_{A}(1)$ axial vector anomaly ~\cite{weinberg}.
The light pseudoscalar mesons ($\pi$, $K$, $\eta$) are the Nambu-Goldstone bosons associated with the spontaneous breaking of the QCD chiral symmetry. Introducing the current quark masses these mesons together with the heavier $\eta^\prime$(958) meson show a mass spectrum which is believed to be explained by the explicit flavor $SU(3)$ breaking and the axial $U_{A}(1)$ anomaly. Recently, there have been several important developments in the study of the spontaneous breaking of chiral symmetry and its partial restoration at finite density~\cite{costa,jido}. An indirect evidence has been claimed for a dropping $\eta^\prime$ mass in the hot and dense matter formed in ultrarelativistic heavy-ion collisions at RHIC energies \cite{Csoergo}.\\
An experimental approach to learn about the $\eta^\prime N$ interaction is the study of $\eta^\prime $ photoproduction off nuclei which provides information on in-medium properties of the $\eta^\prime$-meson. The in-medium width of the $\eta^\prime $-meson can be extracted from the attenuation of the $\eta^\prime$-meson flux deduced from a measurement of the transparency ratio for a number of nuclei. Unless when removed by inelastic channels the $\eta^\prime$-meson will decay outside of the nucleus because of its long lifetime and thus its in-medium mass is not accessible experimentally. The in-medium width provides information on the strength of the $\eta^\prime N$ interaction, as it is studied in \cite{oset_ramos}, and it will be instructive to compare this result with in-medium widths obtained for other mesons. Furthermore, knowledge of the $\eta^\prime$ in-medium width is important for the feasibility of observing $\eta^\prime$ - nucleus bound systems theoretically predicted in some models \cite{Nagahiro}. Until the last  two years $\eta^\prime$ photoproduction was not much explored, but with new generation experiments results on differential and total cross sections of $\eta^\prime$ photoproduction on the proton ~\cite{crede} and the deuteron ~\cite{Igal} have now become available. \\

\section{Data Analysis}
\subsection{\it Experiment}
\begin{figure*} 
 \resizebox{0.9\textwidth}{!}{
     \includegraphics[height=.35\textheight]{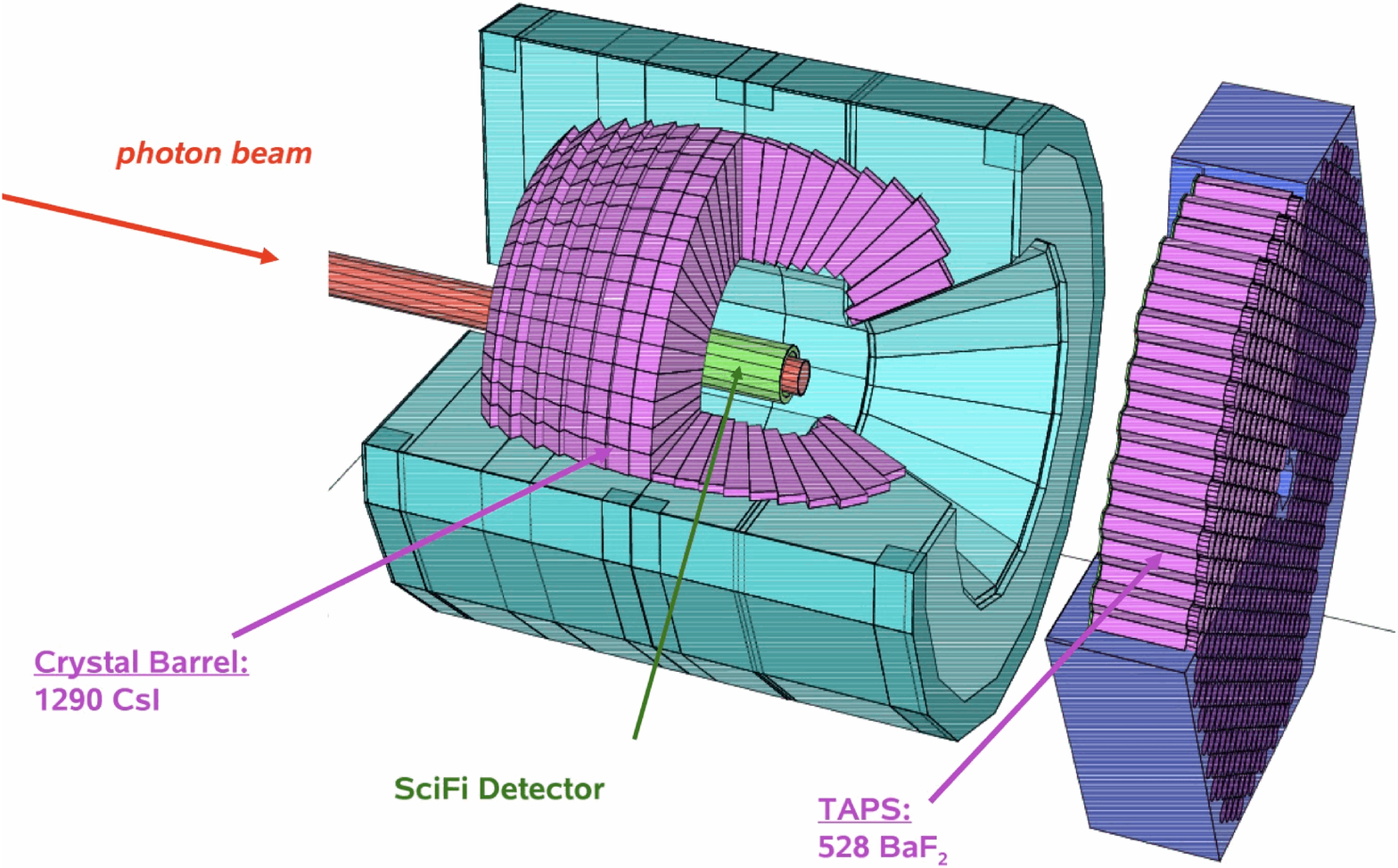}
      \includegraphics[height=.35\textheight]{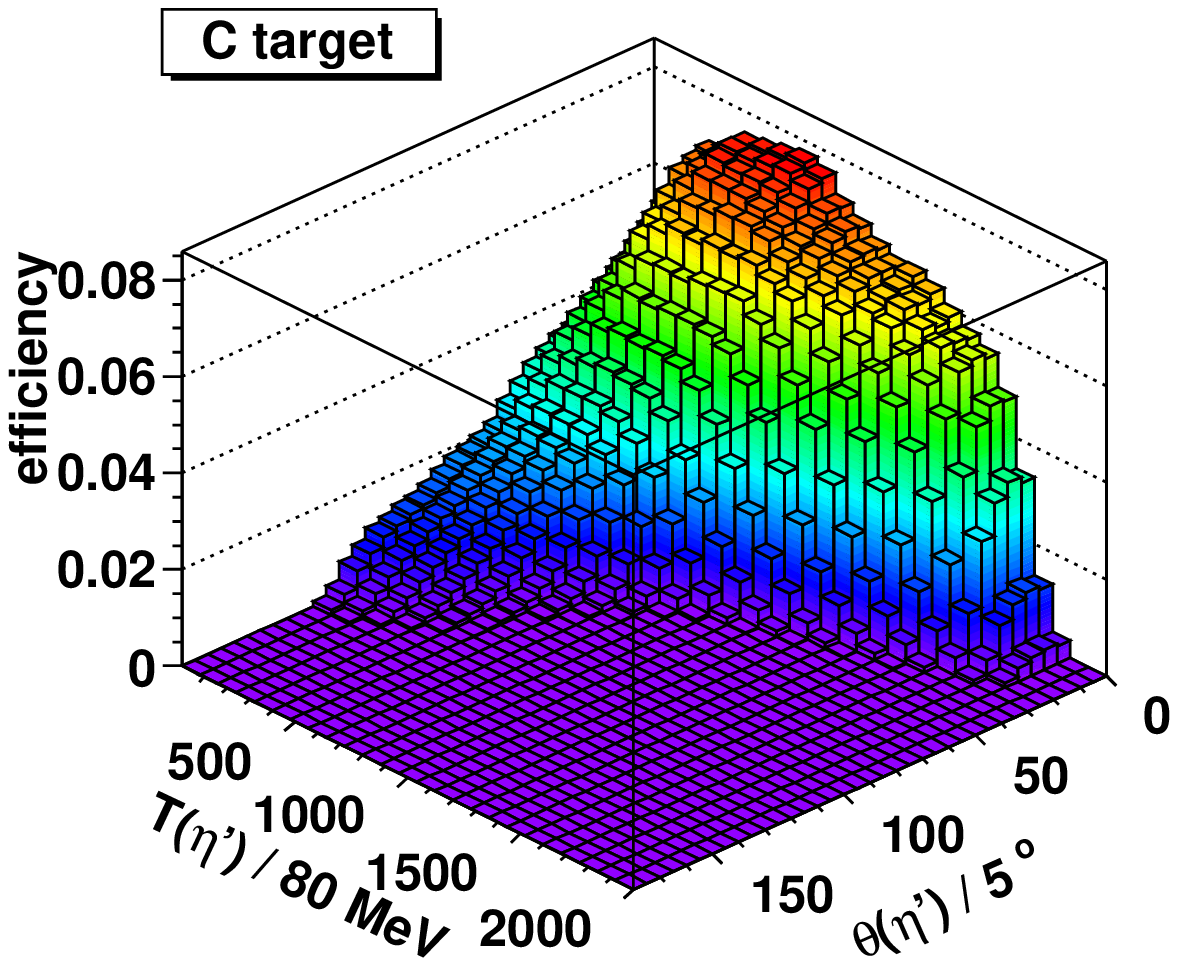}}
 \caption{(Color online) Left: Sketch of the  CB/TAPS setup. The tagged photons impinge on the nuclear target in the center of the Crystal Barrel detector. The   TAPS detector at a distance of 1.18 m from the target serves as a forward wall of the Crystal Barrel. The combined detector system provides photon detection capability over almost the full solid angle. Charged particles leaving the target are identified in the inner scintillating-fiber detector and in the plastic scintillators in front of each BaF$_2$ crystal in TAPS. Right: Detector acceptance for the $ \eta^{'}\rightarrow \pi^{0} \pi^{0} \eta$ decay as a function of the kinetic energy (T) and  the polar angle ($\theta$)  for an  incident photon energy range of 1200 to 2200 MeV. The simulation is for a carbon ($C$) target, taking the trigger conditions into account.} \label{fig:exp} 
\end{figure*}

The experiment has been performed at the ELSA facility in Bonn ~\cite{Husmann,Hillert} using the Crystal Barrel(CB) ~\cite{aker92}  and TAPS~\cite{Novotny1}  detector system (Fig. ~\ref{fig:exp} left). The combined Crystal Barrel/TAPS detector covered 99\% of the full 4$\pi$ solid angle. The high granularity of the system makes it very well 
suited for the detection of multi-photon final states.  Tagged photon beams of energy up to 2.6 GeV were produced via bremsstrahlung and impinged on a solid target. For the measurements targets of ${}^{12}\textrm{C}, {}^{40}\textrm{Ca}, {}^{93}\textrm{Nb}$ and ${}^{208}\textrm{Pb}$ were used. A more detailed description of the detector setup and the running conditions have been given in ~\cite{nanova, Elsner}.

\subsection{\it Acceptance}
The detector acceptance was determined by Monte Carlo simulations using the GEANT3 package, including all features of the detector system, trigger conditions and all cuts for particle indentification. To avoid further uncertainties due to reaction kinematics and final state interactions, the $\eta^\prime$-meson detection efficiency was simulated as a function  of the kinetic energy and the polar angle - $\epsilon(T_{\eta^\prime}, \theta_{\eta^\prime})$. Typical efficiencies are 5-7\% and also slightly different for the different targets. The detection efficiency for $\eta^\prime$ on the $C$ target, taking the trigger conditions into account, is shown in Fig. \ref {fig:exp} (right). Experimental data are efficiency corrected event-by-event with this acceptance as described in \cite{Igal}.\\

\subsection{\it Reconstruction of the $\eta^\prime$-meson}
\begin{figure*}
 \begin{center}
  \resizebox{0.5\textwidth}{!}{%
  \includegraphics[width=0.35\textwidth]{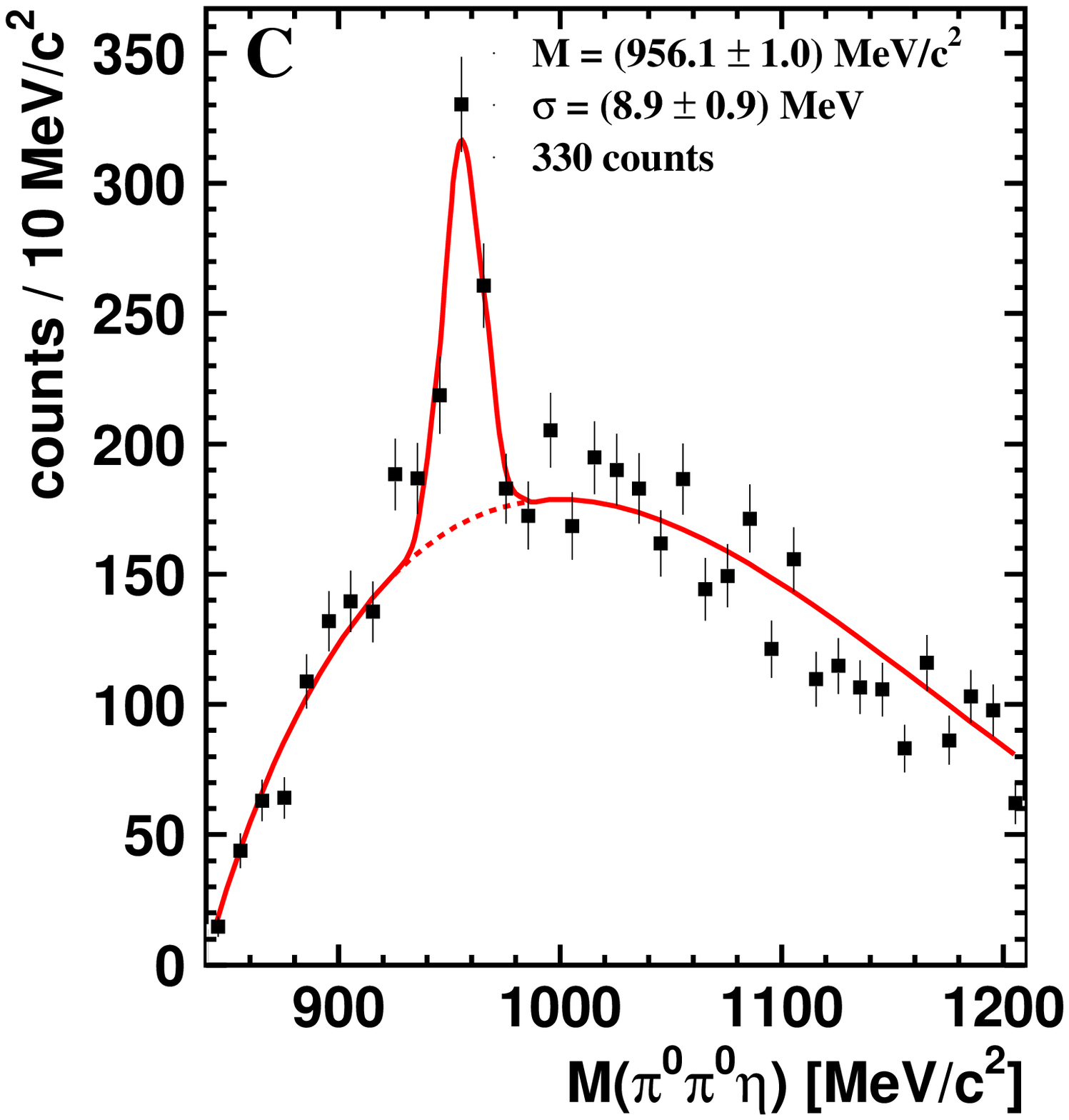}\includegraphics[width=0.35\textwidth]{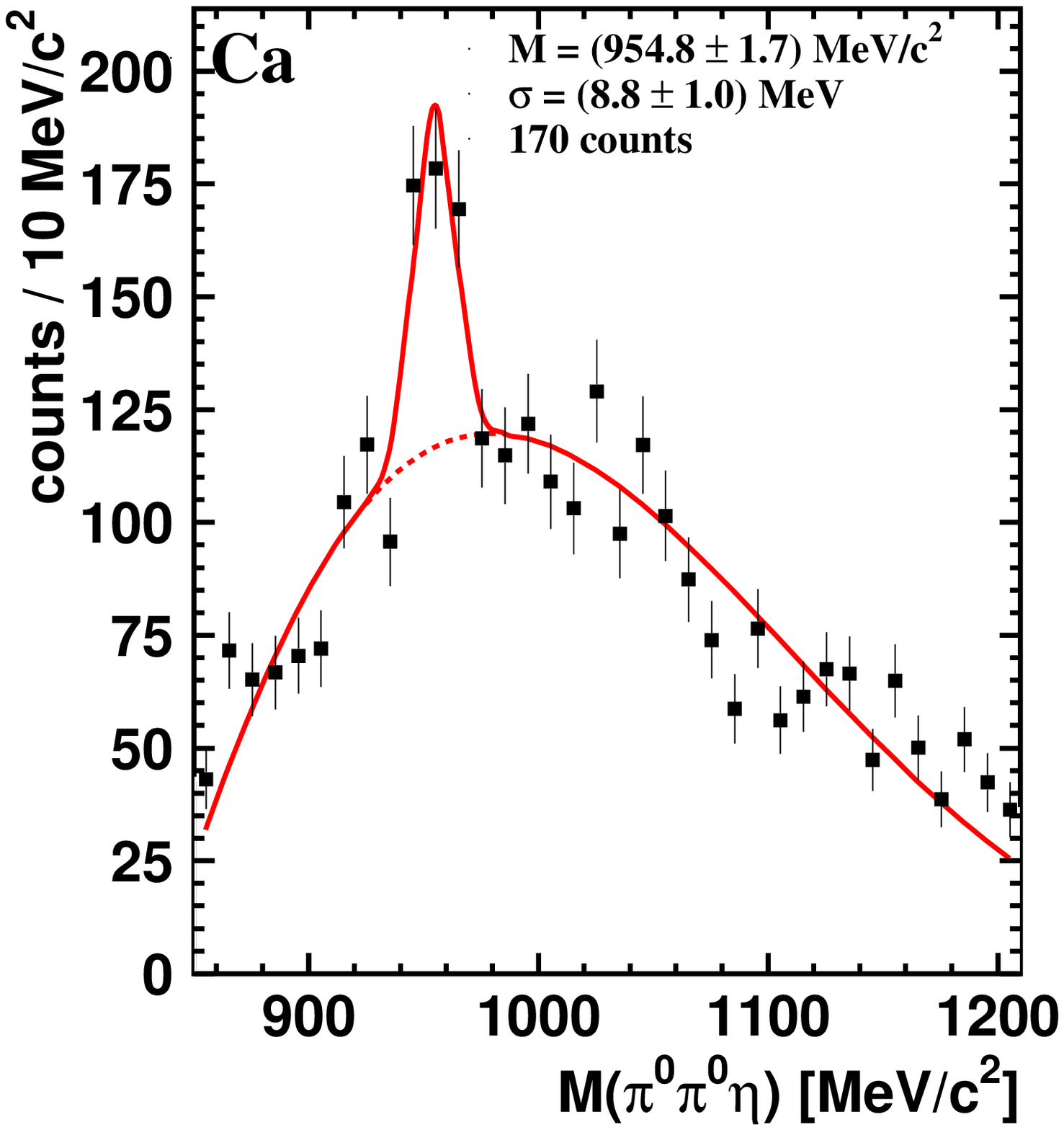}}
   \resizebox{0.5\textwidth}{!}{%
   \includegraphics[width=0.35\textwidth]{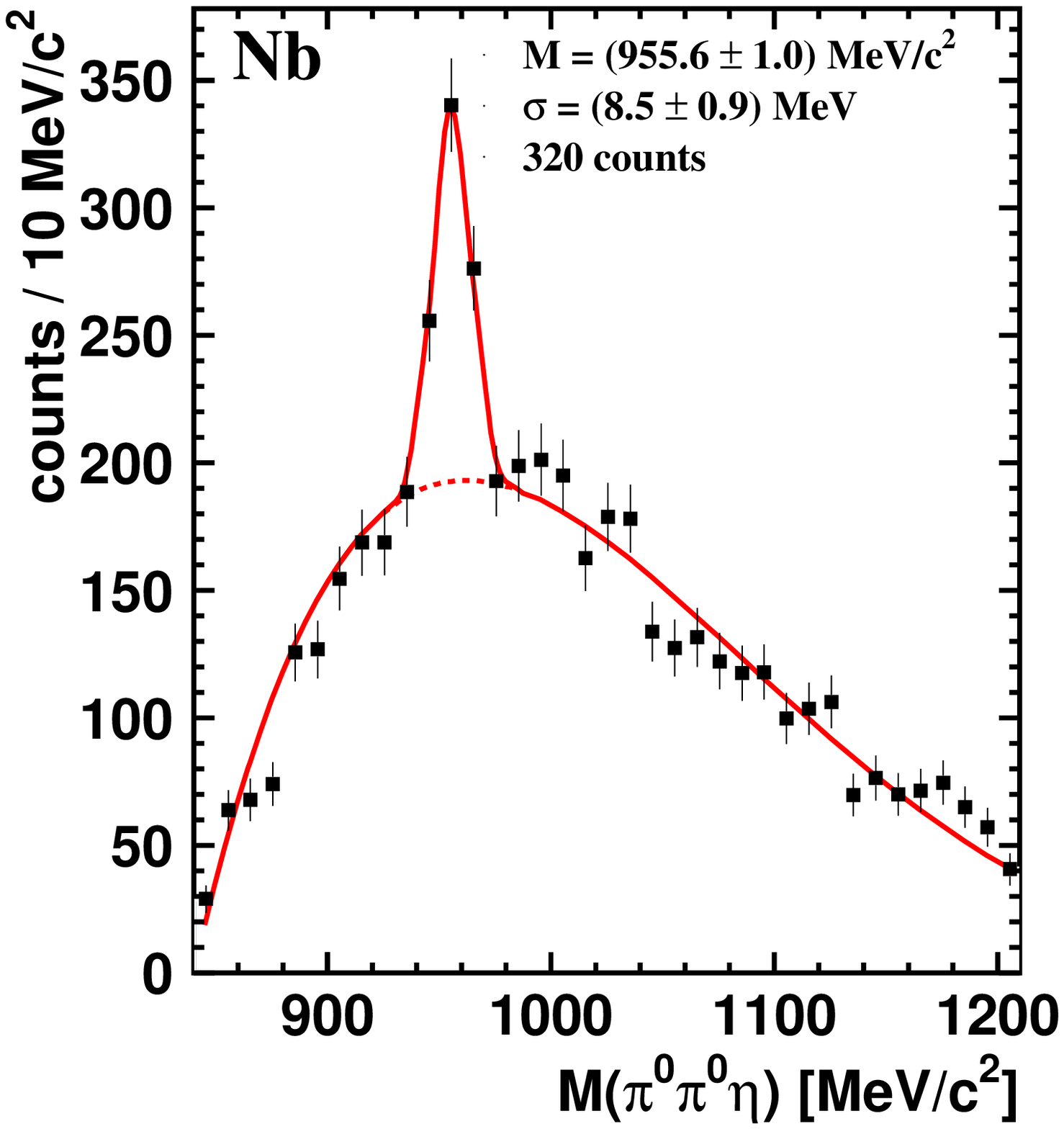}\includegraphics[width=0.35\textwidth]{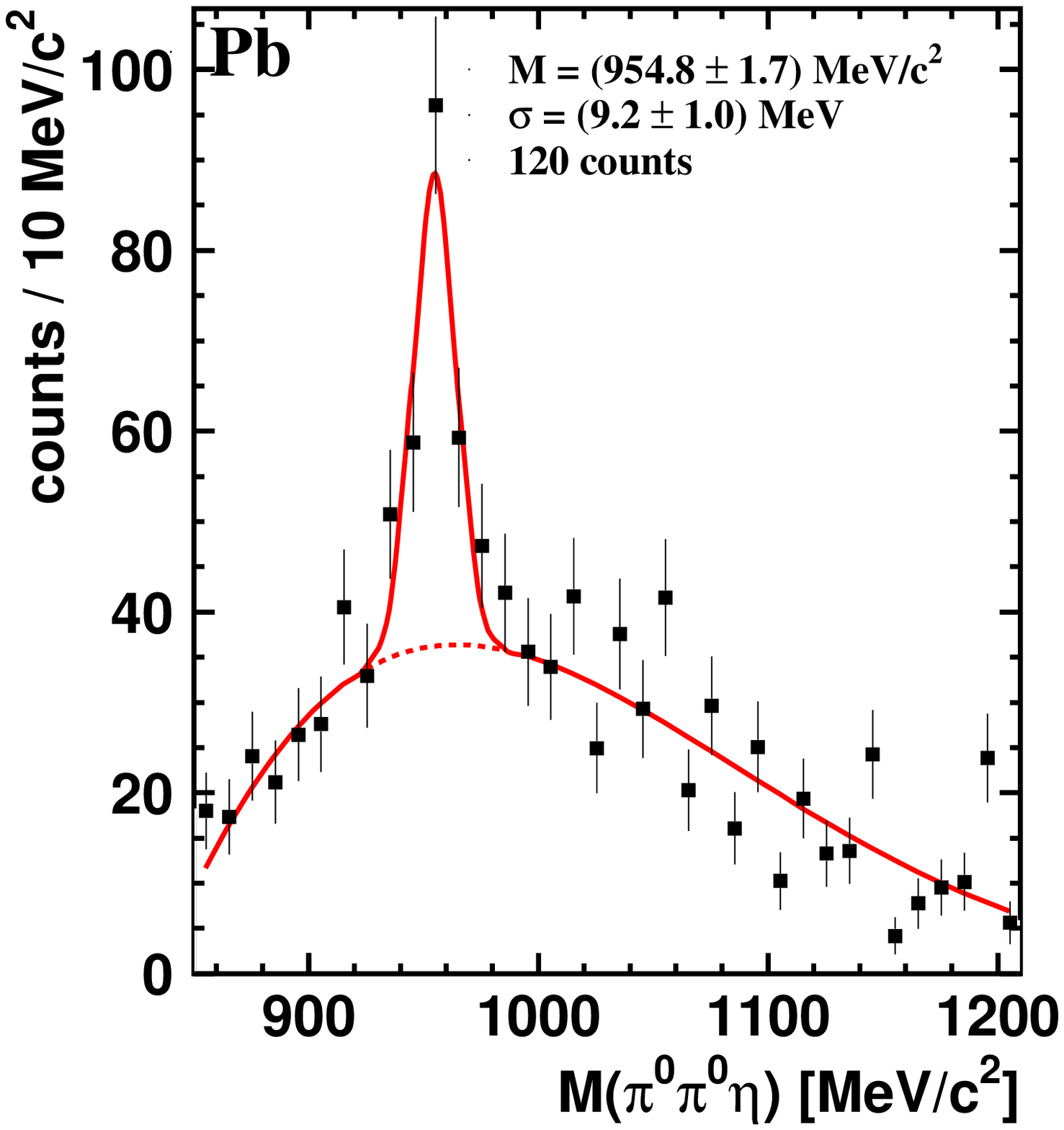}}
\caption{$\pi^{0}\pi^{0}\eta$  invariant mass spectrum for ${}^{12}\textrm{C}, {}^{40}\textrm{Ca}, {}^{93}\textrm{Nb}$ and ${}^{208}\textrm{Pb}$ targets in the incident photon energy range 1200 - 2200 MeV. The distributions have not been corrected for the detector acceptance. The solid curve is a fit to the spectrum. \label{fig:invmass}}
 \end{center}
\end{figure*}

The $\eta^\prime$-mesons were identified via the $\eta^\prime \rightarrow \pi^{0} \pi^{0} \eta \rightarrow 6 \gamma$ decay channel, which has a branching ratio of 8\%. For the reconstruction of the $\eta^\prime$-mesons only events with 6 or 7 neutral hits have been selected. The competing channel with the same final state, namely $\eta \rightarrow \pi^{0} \pi^{0} \pi^{0} \rightarrow 6 \gamma$,  has been reconstructed and the corresponding events have been rejected from the further analysis. In addition only events were kept with at least one combination of the 6 photons to two photon pairs with invariant masses between 110 and 160 MeV ($\pi^{0}$) and one pair between 500 and 600 MeV ($\eta$). The $\pi^{0} \pi^{0} \eta$ invariant mass distributions for the different solid targets are shown in Fig.~\ref{fig:invmass}. The spectra were fitted with a Gaussian and a background function $f(m) = a \cdot (m-m_1)^{b} \cdot (m-m_2)^{c} $.  Alternatively, the background shape was also fitted with a polynomial.The resulting cross sections are used to calculate the transparency ratio of the $\eta^\prime$-meson for a given nucleus $A$ from the formula~(\ref{eq:trans}), normalized to the carbon data:
\begin{equation}
T_A=\frac{12 \cdot \sigma_{\gamma A\to \eta^\prime A^\prime}}{A \cdot
\sigma_{\gamma C\to \eta^\prime C} } \ .
\label{eq:trans}
\end{equation} 

$T_{A}$ describes the loss of flux of $\eta^\prime$-mesons in nuclei via inelastic processes like: $\eta' N \rightarrow \pi^{0} N$. To avoid systematic uncertainties due to unknown secondary production processes, the transparency ratio has been normalized to a light target with equal numbers of protons and neutrons (C) and not to the cross section on the nucleon.

\section{Results and Discussion}
\begin{figure}[tb]
  \begin{center}
\begin{minipage}[t]{8 cm}
\epsfig{file=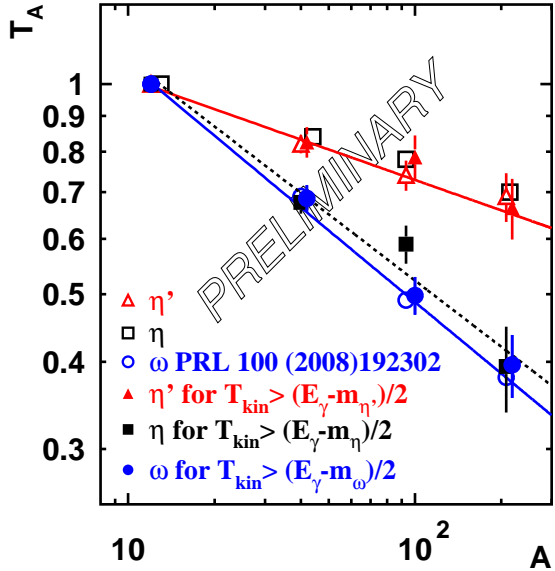,scale=0.5}
\end{minipage}
\begin{minipage}[t]{16.5 cm}
\caption{Transparency ratio for different mesons - $\eta$(squares), $\eta^\prime$(triangels) and $\omega$(circles) as a function of the nuclear mass number $A$.  The transparency ratio with a cut on the kinetic energy for the respective mesons is shown with full symbols. The incident photon energy is in the range 1500 to 2200~MeV. The solid lines are fits to the data. See text for more details. \label{fig:ta}} 
\end{minipage}
\end{center}
\end{figure}

The transparency ratio has been extracted as defined in Eq.~\ref{eq:trans} and is shown in Fig.~\ref{fig:ta} for the $\eta^\prime$-meson (red triangles) compared to the transparency ratio of the $\omega$-meson (blue circles) measured by~\cite{kottu} and the transparency ratio of the $\eta$-meson (black squares) deduced from ~\cite{thierry}. The solid lines are fits to the data points, yielding slope parameters of -0.14 and -0.34 for $\eta^\prime$, 0.34 for $\omega$, respectively.   As it can be seen from the figure, $\eta^\prime$-mesons are only weakly absorbed via inelastic channels as compared to the $\omega$-meson. 

The contribution of secondary production processes like $\pi N \rightarrow  \eta^{'} N$ could increase the number of observed $\eta^\prime$-mesons and thus distort the transparency ratio. The reproduced mesons via such processes should have relativ low kinetic energy. Applying a cut on the kinetic energy of the meson it is possible to study the effect of the secondary production processes on the transparency ratio. In the Fig.~\ref{fig:ta} the data are shown for the full kinetic energy range  of recoiling mesons (open symbols) as well as for the fraction of high energy mesons (full symbols) selected by the constraint

\begin{equation}
T_{kin} \ge (E_{\gamma} -m)/2. \label{cut}
\end{equation}
 Here, $E_{\gamma} $ is the incoming photon energy and $T_{kin} $ and $m$ are the kinetic energy and the mass of the meson, respectively. As discussed in \cite{thierry}, this cut suppresses meson production in secondary reactions by selecting mesons with higher kinetic energy. Fig.~\ref{fig:ta} shows that within errors this cut does not change the experimentally observed transparency ratios for the $\omega$-meson and $\eta^\prime$-meson while there is a significant difference for the $\eta$-meson. For the latter,  the transparency ratio has changed dramatically and shows a slope of -0.32 (dashed line in Fig.~\ref{fig:ta}) which is quite different from the previous one (slope -0.14 - the empty black squares in Fig.~\ref{fig:ta}).
 For the $\eta$-meson secondary production processes appear to be more likely in the relevant photon energy range because of the larger available phase space due to its lower mass (547~MeV/c$^2$) compared to the $\omega$ (782~MeV/c$^2$) and $\eta^\prime$ (958~MeV/c$^2$) meson. Cross sections for pion-induced reactions favor secondary production processes in case of the $\eta$-meson: 3 mb at $p_{\pi} \approx$ 750 MeV/c in comparison to 2.5 mb at $p_{\pi} \approx $ 1.3 GeV/c for the $\omega$-meson and 0.1 mb at $p_{\pi} \approx$ 1.5 GeV/c for the $\eta^\prime$-meson, respectively \cite{Landolt-Boernstein}. In addition, $\eta$-mesons may be slowed down by absorption of an initially produced $\eta$-meson which is then re-emitted with lower kinetic energy from a S$_{11}$(1535) resonance. According to Fig.~\ref{fig:ta}  the $\eta^\prime$-meson shows a much weaker attenuation in normal nuclear matter than the $\omega$ and $\eta$-meson which exhibit a similarly strong absorption after suppressing secondary production effects in case of the $\eta$-meson.\\
Since the transparency ratios for $\eta^\prime$-mesons measured for several nuclei deviate sufficiently from unity (Fig.~\ref{fig:ta}) an extraction of the  $\eta^\prime $ width in the nuclear medium is possible using the Glauber model in the high energy eikonal approximation~\cite{glauber}. It was demonstrated in~\cite{muelich} that the Glauber model provides a quite reliable tool to extract the $\omega N$ inelastic cross section. According to~\cite{muelich} the transparency ratio can be expressed by\\
\begin{equation}
T_A=\frac{\pi R^{2}}{A \sigma_{\eta^\prime N}} \{1+(\frac{\lambda}{R}) exp\left[-2\frac{R}{\lambda} \right ] +\frac{1}{2}(\frac{\lambda}{R})^{2} (exp \left [ -2\frac{R}{\lambda} \right ]-1) \} \label{MM_fit}
\end{equation}
where $\lambda= (\rho_{0} \cdot \sigma_{\eta^\prime N})^{-1} $ is the mean free path of the $\eta^\prime$-meson in a nucleus with density $\rho_{0}$=0.17 fm$^{-3}$ and radius $R=r_{0} \cdot A^{1/3}$, with $r_{0}$ = 1.143 $fm$.  Fitting this expression to the $\eta^\prime$ transparency ratio data an in-medium $\eta^\prime$N inelastic cross section of $\sigma_{\eta^\prime N}$ = (11.5$\pm$1.6) mb is derived~\cite{karoly}. The in-medium width is then obtained from
\begin{equation}
\Gamma = \hbar c \cdot \rho \cdot \sigma_{inel} \cdot \beta, \label{eq:Gamma-sigma}
\end{equation}
with
\begin{equation}
\beta = \frac{p_{\eta^\prime}}{E_{\eta^\prime}}
\end{equation}
in the laboratory, where the average $\eta^\prime$ recoil momentum is 1.05 GeV/c. The resulting in-medium width of the $\eta^\prime$-meson is $\Gamma \approx$ 25 MeV.
  
More precise information on the in-medium width of the $\eta^\prime$-meson will become available by comparing the experimental data with calculations based on recent studies of the $\eta^\prime N$ interaction~\cite{oset_ramos} including the measurement of the $\eta^\prime$ photoproduction off the proton~\cite{crede}.  
 \section{Conclusion}
 The transparency ratios for $\eta^\prime$-mesons have been measured for several nuclei. A comparison of these results with corresponding measurements for $\omega$- and $\eta$-mesons demonstrates the relatively weak interaction of the $\eta^\prime$-meson with nuclear matter. Using the Glauber model approximation an inelastic cross section for $\eta^\prime N$ interaction of $\sigma_{\eta^\prime N}$=(11.5$\pm$1.6) mb is derived, corresponding to an in-medium width of the $\eta^\prime$-meson of $\Gamma \approx$ 25 MeV. This relatively small width encourages the search for $\eta^\prime$-nuclear bound states.\\
 This work was supported by the DPG through SFB/TR16.

\end{document}